\documentclass[journal=jctcce,manuscript=article,layout=twocolumn]{achemso}

\usepackage[colorlinks,linkcolor=blue,citecolor=blue,urlcolor=black,bookmarks=false,hypertexnames=true]{hyperref} 

\usepackage[version=3]{mhchem}
\usepackage{amssymb}
\usepackage{color}
\usepackage{braket}
\usepackage{xspace}
\usepackage{cleveref}
\usepackage{graphicx}
\usepackage{subfigure}
\usepackage{threeparttable}
\usepackage{textcomp}
\usepackage{setspace}
\usepackage{caption}
\usepackage{comment}

\usepackage{array}
\newcolumntype{L}[1]{>{\raggedright\let\newline\\\arraybackslash\hspace{0pt}}m{#1}}
\newcolumntype{C}[1]{>{\centering\let\newline\\\arraybackslash\hspace{0pt}}m{#1}}
\newcolumntype{R}[1]{>{\raggedleft\let\newline\\\arraybackslash\hspace{0pt}}m{#1}}

\makeatletter
\let\l@addto@macro\relax
\makeatother
\usepackage[fontsize=11pt]{scrextend}

\let\oldmaketitle\maketitle
\let\maketitle\relax

\newcommand{\e}[1]{\ensuremath{\varepsilon_{#1}}}

\newcommand{\kc}{\textbf{k}}

\newcommand{\angstrom}{\mbox{\normalfont\AA}\xspace}
\newcommand*{\maxe}{$\Delta_{\mathrm{MAX}}$\xspace}
\newcommand*{\mae}{\ensuremath{\Delta_{\mathrm{MAE}}}\xspace}
\newcommand*{\std}{\ensuremath{\Delta_{\mathrm{STD}}}\xspace}

\crefname{figure}{Figure}{Figures}
\crefname{table}{Table}{Tables}
\crefname{equation}{Eq.}{Eqs.}
\crefname{section}{Section}{Sections}
\crefname{subsection}{Section}{Sections}

\author{Samragni Banerjee}
\affiliation{%
     Department of Chemistry and Biochemistry,
     The Ohio State University,
     Columbus, Ohio 43210, United States
}
 \author{Alexander Yu.\ Sokolov}
 \email{sokolov.8@osu.edu}
 \affiliation{%
     Department of Chemistry and Biochemistry,
     The Ohio State University,
     Columbus, Ohio 43210, United States
 }

\begin{tocentry}
\includegraphics{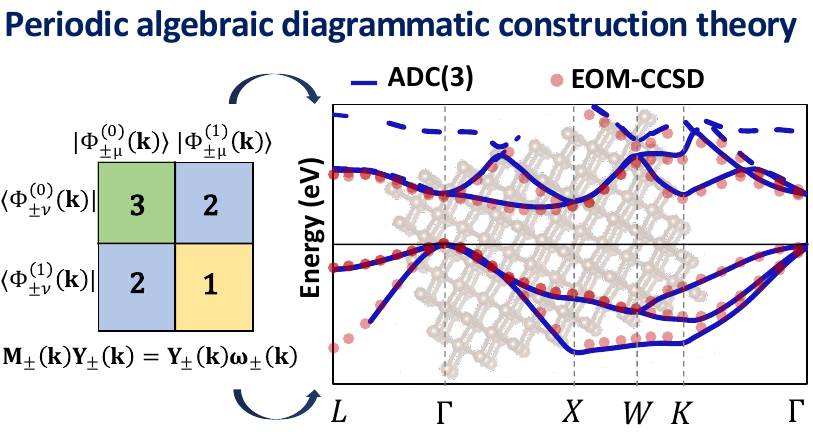}
\end{tocentry}

\title{{\color{blue}Non-Dyson Algebraic Diagrammatic Construction Theory for Charged Excitations in Solids}}

\begin{document}

\newcommand*{\abstractext}{
We present the first implementation and applications of non-Dyson algebraic diagrammatic construction theory for charged excitations in three-dimensional periodic solids (EA/IP-ADC).
The EA/IP-ADC approach has a computational cost similar to the ground-state M\o ller--Plesset perturbation theory, enabling efficient calculations of a variety of crystalline excited-state properties (e.g., band structure, band gap, density of states) sampled in the Brillouin zone. 
We use EA/IP-ADC to compute the quasiparticle band structures and band gaps of several materials (from large-gap atomic and ionic solids to small-gap semiconductors) and analyze the errors of EA/IP-ADC approximations up to the third order in perturbation theory.
Our work also reports the first-ever calculations of ground-state properties (equation-of-state and lattice constants) of three-dimensional crystalline systems using a periodic implementation of third-order M\o ller--Plesset perturbation theory (MP3). 
\vspace{0.25cm}
}

\twocolumn[
\begin{@twocolumnfalse}
\oldmaketitle
\vspace{-0.75cm}
\begin{abstract}
\abstractext
\end{abstract}
\end{@twocolumnfalse}
]

\section{Introduction}
\label{sec:intro}

Accurate simulations of ground- and excited-state electronic structure of solids are crucial for the development of new materials with desired properties.\cite{Scuseria:2021p118}
Periodic (usually, plane-wave) density functional theory (DFT)\cite{Sham:1965pA1133,Hafner:2008p2044,Li:2013p2950}  is a de facto standard theoretical approach for the ground-state calculations of crystalline systems due to its relatively low computational cost and ability to incorporate electron correlation effects in the exchange-correlation functional.
To obtain the excited-state properties (e.g.,\@ band structure, fundamental and optical band gaps), DFT is usually combined with Green's function-based many-body perturbation theory, such as the GW approximation\cite{Hedin:1965p796,Faleev:2004p126406,vanSchilfgaarde:2006p226402,Neaton:2006p216405,Samsonidze:2011p186404,vanSetten:2013p232,Reining:2017pe1344} and Bethe--Salpeter equation (BSE).\cite{Albrecht:1998p189,Hanke:1980p4656,Sham:1966p708,Strinati:1984p5718,Rohlfing:2000p4927} 
Although these DFT-based approaches provide important insights about the electronic structure of solids,\cite{Faleev:2004p126406,Li:2005p193102,Aryasetiawan:1995p3221,Vinson:2012p195135,Varrassi:2021p074601,Thimsen:2009p2014} their accuracy can be affected by the self-interaction error and the exchange-correlation functional approximations inherent in DFT,\cite{Perdew:2001p1} making the assessment of errors and systematic improvement of these methods difficult to achieve.\cite{Borlido:2020p96,Reining:2017pe1344} 

Recently, there has been a lot of progress in developing new, systematically improvable methods for solids by adapting the wavefunction-based theories from molecular quantum chemistry to periodic simulations of crystalline systems.
The ground-state properties, such as the lattice parameters and cohesive energies, have been calculated using periodic implementations of second-order M\o ller--Plesset perturbation theory (MP2),\cite{Sun:1996p8553,Ayala:2001p9698,Marsman:2009p184103,Gruneis:2010p074107,Gruneis:2011p2780,Usvyat:2013p194101,Shepherd:2012p035111} coupled cluster theory (CC) with up to triple excitations,\cite{Zhang:2019p123,Hirata:2004p2581,Katagiri:2005p224901,Gruneis:2011p2780,Booth:2013p365,Liao2016:p141102,Irmler:2021p234103,Gruber:2018p021043}
and full configuration interaction quantum Monte Carlo.\cite{Shepherd:2012p081103,Booth:2013p365}
To compute the crystalline band structures and densities of states, equation-of-motion (EOM-CC)\cite{Katagiri:2005p224901,McClain:2017p1209,Gao:2020p165138,Gallo:2021p064106} and Green's function (GF-CC)\cite{Furukawa:2018p204109,Yeh:2021p155158,Zhu:2021p021006,Shee:2022p664} formulations of coupled cluster theory have been also developed. 
However, due to the high computational cost of CC, EOM-CC, and GF-CC methods, their applications to solids are far from routine. 
Introducing approximations using the local pair natural orbitals\cite{Dittmer:2019p9303} or partitioned EOM-MP2 approach\cite{Lange:2021p081101} has been explored to lower the computational cost.
Despite these efforts, the development of accurate yet computationally affordable theoretical methods for the reliable simulations of crystalline systems is still an ongoing challenge.

In this work, we explore the application of algebraic diagrammatic construction theory (ADC) to charged excited states of solids.
In molecular quantum chemistry, ADC\cite{Schirmer:1982p2395,Schirmer:1983p1237,Schirmer:1991p4647,Mertins:1996p2140,Schirmer:2004p11449,Dreuw:2014p82} is an attractive alternative to EOM- and GF-CC, offering a hierarchy of efficient and systematically improvable approximations with computational cost similar to M\o ller--Plesset perturbation theory.
For charged excitations of molecules, two flavors of ADC have been developed: (i) the Dyson framework,\cite{Schirmer:1983p1237,Angonoa:1987p6789,Santra:2009p013002,Averbukh:2011p36} which computes the charged excitation energies and spectra by perturbatively approximating the self-energy in the Dyson equation, and (ii) the non-Dyson framework,\cite{Schirmer:1998p4734,Trofimov:2005p144115,Schirmer:2001p10621,Thiel:2003p2088,Dempwolff:2019p064108,Banerjee:2019p224112,Dempwolff:2020p024113,Banerjee:2021p074105,Dempwolff:2021p104117} which directly approximates the one-particle Green's function bypassing the solution of the Dyson equation. 
Only the Dyson framework has been extended to periodic systems with limited benchmark applications to ionic crystals and one-dimensional molecular chains.\cite{Buth:2005p1,Buth:2006,Bezugly:2008p012006}  
However, the non-Dyson ADC has several important advantages over the Dyson framework.
Similar to EOM- and GF-CC, the non-Dyson ADC methods are derived by decoupling the forward and backward components of the one-particle Green's function allowing for independent calculations of ionized and electron-attached states, while this coupling has to be considered in Dyson ADC leading to a computational overhead.  
For this reason, the non-Dyson ADC framework is computationally more efficient and is nowadays predominantly used for the ADC calculations of charged excitations in molecules. 

Here, we extend the non-Dyson formulation of ADC to simulations of electron-attached (EA) and ionized (IP) states in periodic three-dimensional systems.
We present the first derivation and implementation of EA/IP-ADC with periodic boundary conditions that takes advantage of crystalline translational symmetry (\cref{sec:theory,sec:implementation}). 
Further, we apply the EA/IP-ADC methods to compute the band structures and band gaps of several materials, from large-gap atomic and ionic solids to small-gap semiconductors (\cref{sec:results:band_structures,sec:results:band_gaps}).
Our work also reports the first-ever periodic calculations of three-dimensional materials using third-order M\o ller--Plesset perturbation theory (\cref{sec:results:equation_of_state}).
Our results highlight the strengths and weaknesses of low-order periodic EA/IP-ADC approximations, establishing a foundation  for their future development.

\section{Theory}
\label{sec:theory}
\subsection{Non-Dyson ADC for charged excitations of periodic systems}
\label{sec:theory:overview_adc}

We begin by formulating the framework of non-Dyson algebraic diagrammatic construction (ADC) theory \cite{Schirmer:1982p2395,Schirmer:1998p4734,Trofimov:2005p144115,Schirmer:2001p10621,Thiel:2003p2088,Dempwolff:2019p064108,Banerjee:2019p224112,Dempwolff:2020p024113,Banerjee:2021p074105,Dempwolff:2021p104117} for electron attachment and ionization of periodic systems.
In ADC, these charged excitations are described using the retarded single-particle Green's function (1-GF).\cite{Fetter:1971quantum,Dickhoff:2008many}
The exact 1-GF of a periodic system can be written in the reciprocal $k$-space as
\begin{align}
G_{pq}(\omega,\kc) &= G_{pq}^{+}(\omega,\kc) + G_{pq}^{-}(\omega,\kc)\notag\\
               &=\langle\Psi_{0}^{N}|a_{p\kc} (\omega - H + E_{0}^{N})^{-1}a_{q\kc}^{\dagger}|\Psi_{0}^{N}\rangle\notag\\
               &+ \langle\Psi_{0}^{N}|a_{q\kc}^{\dagger} (\omega + H - E_{0}^{N})^{-1}a_{p\kc}|\Psi_{0}^{N}\rangle \ , \label{eq:G_pq}
\end{align}
where $G_{pq}^{+}(\omega,\kc)$  and $G_{pq}^{-}(\omega,\kc)$ are the forward and backward components of 1-GF describing the electron attachment (EA) and ionization (IP) processes, respectively, while $|\Psi_{0}^{N}\big\rangle$ and  $E_{0}^{N}$ are the exact ground-state wavefunction and energy of an $N$-electron system with electronic Hamiltonian $H$.
The electron attachment and ionization are described using the fermionic creation ($a_{p\kc}^{\dag}$) and annihilation ($a_{p\kc}$) operators.
The 1-GF in \cref{eq:G_pq} is a function of two variables: the frequency $\omega$ and the crystal momentum vector $\kc$.
The frequency is, in general, a complex variable $\omega = \Re(\omega) + i \Im(\omega)$.
The real part of $\omega$ corresponds to physical frequencies that define many important properties of a periodic system (e.g., band gap, density of states), while the complex part is used to work away from the real axis (e.g., to introduce a small broadening in the density of states).
Due to the translational symmetry of $H$ and crystal momentum conservation, 1-GF depends only on one crystal momentum $\kc$.

The exact 1-GF in \cref{eq:G_pq} can be expressed using a more compact tensor notation in the spectral (or Lehmann) representation \cite{Kobe:1962p448}
\begin{align}
    \textbf{G}_{\pm}(\omega,\kc) = \mathbf{\tilde{X}}_{\pm}(\kc)(\omega\textbf{1} - \boldsymbol{\tilde{\Omega}}_{\pm}(\kc))^{-1}\mathbf{\tilde{X}}_{\pm}^{\dagger}(\kc) \ , \label{eq:G_pq_matrix}
\end{align}
where $\boldsymbol{\tilde{\Omega}}_{\pm}(\kc)$ are the $\kc$-dependent diagonal matrices of vertical attachment ($\omega_{+n}(\kc) = E_{n}^{N+1}(\kc) - E_{0}^{N}$) and ionization ($\omega_{-n}(\kc) = E_{0}^{N} - E_{n}^{N - 1}(\kc)$) energies and $\mathbf{\tilde{X}}_{\pm}(\kc)$ are the matrices of spectroscopic amplitudes $\tilde{X}_{+pn}(\kc) = \langle\Psi_{0}^{N}|a_{p\kc}|\Psi_{n}^{N+1}(\kc)\rangle$ and $\tilde{X}_{-qn}(\kc) = \langle\Psi_{0}^{N}|a_{q\kc}^{\dagger}|\Psi_{n}^{N-1}(\kc)\rangle$. 
The charged excitation energies $\boldsymbol{\tilde{\Omega}}_{\pm}(\kc)$ provide information about the crystal band structure and can be used to calculate a fundamental band gap.
The spectroscopic amplitudes define the 1-GF pole strengths that can be interpreted as the intensities in photoelectron spectroscopy experiments. 

In non-Dyson ADC, the EA and IP components of 1-GF are approximated using perturbation theory up to a particular order $n$:
\begin{align} 
	\textbf{G$_{\pm}(\omega,\kc)$} 
	&\approx \textbf{G$_{\pm}^{(0)}(\omega,\kc)$} +\textbf{G$_{\pm}^{(1)}(\omega,\kc)$} + ... \notag\\
	&+ \textbf{G$_{\pm}^{(n)}(\omega,\kc)$} \ . \label{eq:G_n}
\end{align}
In contrast to the periodic formulation of Dyson ADC,\cite{Buth:2005p1,Buth:2006,Bezugly:2008p012006} the expansion in \cref{eq:G_n} is carried out for $\textbf{G}_{+}(\omega,\kc)$ and $\textbf{G}_{-}(\omega,\kc)$ independently, so that the two components do not couple and can be calculated separately. 
In order to define the perturbative series in \cref{eq:G_n}, we assume that the ground-state wavefunction of a periodic system $|\Psi_{0}^{N}\big\rangle$ can be well-approximated by a Slater-determinant wavefunction $\ket{\Phi}$ obtained from a periodic mean-field (Hartree--Fock) calculation. 
The approximate 1-GF is expressed in a non-diagonal tensor form using an approximate basis of ($N\pm 1$)-electron states
\begin{align}
    \textbf{G}_{\pm}(\omega,\kc) = \textbf{T}_{\pm}(\kc)(\omega\textbf{S}_{\pm}(\kc)-\textbf{M}_{\pm}(\kc))^{-1}\textbf{T}_{\pm}^{\dagger}(\kc) \ . \label{eq:G_approx}
\end{align}
In \cref{eq:G_approx}, $\textbf{M}_{\pm}(\kc)$ is the non-diagonal effective Hamiltonian matrix that contains the information about approximate charged excitation energies and $\textbf{T}_{\pm}(\kc)$ is the effective transition moments matrix that can be used to calculate the approximate pole strengths. 
Although, in general, the approximate basis can be non-orthogonal, in this work we assume that we use an orthogonal basis with an overlap matrix $\textbf{S}_{\pm}(\kc)=\textbf{1}$. 
The $n$th-order ADC approximations (EA/IP-ADC($n$)) are defined by evaluating each matrix in \cref{eq:G_approx} up to the perturbation order $n$. 

\cref{eq:G_approx} can be rewritten in a diagonal (or spectral) representation
\begin{align}
	\label{eq:g_adc}
	\mathbf{G}_{\pm}(\omega,\kc) &= \mathbf{X}_{\pm}(\kc) \left(\omega \mathbf{1} - \boldsymbol{\Omega}_{\pm}(\kc)\right)^{-1}  \mathbf{X}_{\pm}(\kc)^\dag 
\end{align}
where $\boldsymbol\Omega_{\pm}(\kc)$  is a matrix of the EA/IP-ADC($n$) charged excitation (or so-called band) energies that are calculated by diagonalizing $\textbf{M}_{\pm}(\kc)$
\begin{align}
	\textbf{M}_{\pm}(\kc)\textbf{Y}_{\pm}(\kc)=\textbf{Y}_{\pm}(\kc)\boldsymbol{\Omega}_{\pm}(\kc) \label{eq:eig}
\end{align}
and $\textbf{X}_{\pm}(\kc)$ is a matrix of the EA/IP-ADC($n$) spectroscopic amplitudes
\begin{align}
    \textbf{X}_{\pm}(\kc)=\textbf{T}_{\pm}(\kc) \textbf{Y}_{\pm}(\kc) \ . \label{eq:spec_amp}
\end{align}
The pole strengths of the corresponding transitions are determined by calculating the spectroscopic factors 
\begin{align}
	\label{eq:spec_factors}
	P_{\pm\mu}(\kc) = \sum_{p} |X_{\pm p\mu}(\kc)|^{2} \ ,
\end{align}
which, together with the band energies $\boldsymbol\Omega_{\pm}(\kc)$, provide information about the band structure and momentum-resolved density of states
\begin{align}
	\label{eq:dos}
	A(\omega,\kc) &= -\frac{1}{\pi} \mathrm{Im} \left[ \mathrm{Tr} \, \mathbf{G}_{\pm}(\omega,\kc) \right] \ .
\end{align}
We note that, in contrast to many other Green's function theories that are widely used for simulations of crystalline systems,\cite{Doniach:1998green,Hedin:1965p796,Faleev:2004p126406,vanSchilfgaarde:2006p226402,Neaton:2006p216405,Samsonidze:2011p186404,vanSetten:2013p232,Reining:2017pe1344} EA/IP-ADC allows to calculate the charged excitation energies and pole strengths directly, without evaluating 1-GF on a grid of frequency points. 

In \cref{sec:theory:periodic_adc_theory}, we provide more details about the derivation of $k$-adapted EA/IP-ADC working equations for periodic systems.

\subsection{Derivation of \kc-conserved working equations for periodic EA/IP-ADC }
\label{sec:theory:periodic_adc_theory}

To derive the equations for periodic EA/IP-ADC, a suitable periodic one-electron basis set needs to be introduced.
In this work, we employ a basis of crystalline molecular orbitals\cite{McClain:2017p1209,Sun:2017p164119} 
\begin{align}
	\psi_{p\kc}(\textbf{r}) = \sum_{\mu}c_{\mu p\kc} \phi_{\mu \kc}(\textbf{r})  \label{eq:crys_MO}
\end{align}
constructed as linear combinations of translation-symmetry-adapted Gaussian atomic orbitals 
\begin{align}
	\phi_{\mu\kc}(\textbf{r}) = \sum_{\textbf{T}}e^{i\kc\cdot\textbf{T}}\chi_{\mu}(\textbf{r} - \textbf{T}) \ , \label{eq:crys_AO} 
\end{align}
where $\chi_{\mu}(\textbf{r} - \textbf{T})$ are the atom-centered Gaussian basis functions, \textbf{T} is a lattice translation vector, and $\kc$ is a crystal momentum vector in the first Brillouin zone.
The expansion coefficients $c_{\mu p\kc}$ are computed by performing a periodic Hartree--Fock calculation in the basis of $\phi_{\mu\kc}(\textbf{r})$.

Having the basis set defined, we express the momentum-conserved crystalline Hamiltonian $H$ in a second-quantized form
\begin{align}
   H &= E_{0} + \sum_{pq}\sum_{\kc}f_{q}^{p}(\kc)\{a_{p\kc}^{\dagger}a_{q\kc}\} \notag \\
   &+  \frac{1}{4}\sum_{pqrs}\sum^{'}_{\substack{\kc_{p}\kc_{q}\\\kc_{r}\kc_{s}}}v^{p\kc_{p}q\kc_{q}}_{r\kc_{r}s\kc_{s}}\{a_{p\kc_{p}}^{\dagger}a_{q\kc_{q}}^{\dagger}a_{s\kc_{s}}a_{r\kc_{r}}\}\label{eq:Ham} \ ,
\end{align}
where $E_{0} = \langle\Phi|H|\Phi\rangle$ is the reference periodic Hartree--Fock energy per unit cell and $f_{q}^{p}$(\kc) is the canonical Fock matrix with the orbital energies $\epsilon_p(\kc)$ in the diagonal elements.\cite{McClain:2017p1209} 
Notation $\{...\}$ indicates that the creation and annihilation operators are normal-ordered with respect to the reference Slater determinant $|\Phi\rangle$. 
The antisymmetrized two-electron integrals per unit cell are defined as $v^{p\kc_{p}q\kc_{q}}_{r\kc_{r}s\kc_{s}} = \langle p\kc_{p}q\kc_{q}|r\kc_{r}s\kc_{s}\rangle - \langle p\kc_{p}q\kc_{q}|s\kc_{s}r\kc_{r}\rangle$, where
\begin{align}
& \langle p\kc_{p}q\kc_{q}|r\kc_{r}s\kc_{s}\rangle = \int_{\Omega} d\textbf{r}_{1} \int d\textbf{r}_{2} \notag \\
 &\times \psi^{*}_{p\kc_{p}}(\textbf{r}_{1}) \psi^{*}_{q\kc_{q}}(\textbf{r}_{2}) \frac{1}{|\textbf{r}_{1} - \textbf{r}_{2}|}  \psi_{r\kc_{r}}(\textbf{r}_{1})\psi_{s\kc_{s}}(\textbf{r}_{2})\label{eq:v_pqrs}
\end{align}
and $\Omega$ is the unit cell volume. 
The primed summation in \cref{eq:Ham} indicates the conservation of crystal momenta, $\kc_{p} + \kc_{q} - \kc_{r} - \kc_{s} = 0$, where $p,q,r,s$ run over all crystalline molecular orbitals. 

To construct the EA/IP-ADC approximations for 1-GF in \cref{eq:G_approx}, we partition the Hamiltonian in \cref{eq:Ham} into a zeroth-order contribution 
\begin{align}
   H^{(0)} = E_{0} + \sum_{p}\sum_{\kc}\epsilon_{p}(\kc)\{a_{p\kc}^{\dagger}a_{p\kc}\} \label{eq:H_0}
\end{align}
and a perturbation $V = H -H^{(0)}$.  
Working equations for the contributions to $\textbf{M}_{\pm}(\kc)$ and $\textbf{T}_{\pm}(\kc)$ at each order in perturbation theory are derived using the formalism of effective Liouvillian theory,\cite{Prasad:1985p1287,Mukherjee:1989p257,Sokolov:2018p204113} which expresses 1-GF using a superoperator\cite{Lowdin:1985p285} of an effective Hamiltonian $\tilde{{H}}=e^{-(T-T^{\dagger})}He^{(T-T^{\dagger})}$ and suitable excitation operator manifolds $h_{\pm\mu}^{\dagger}(\kc)$ that satisfy the vacuum annhilation condition (VAC) \cite{Weiner:1980p1109,Goscinski:1980p385,Prasad:1985p1287,Datta:1993p3632} with respect to the reference wavefunction ($h_{\pm\mu}(\kc) |\Phi \rangle= 0 $). 
The fulfillment of VAC guarantees the decoupling of EA and IP components of 1-GF, which allows for a straightforward derivation and implementation of the non-Dyson EA- and IP-ADC($n$) approximations. 
In the definition of effective Hamiltonian $\tilde{{H}}$, the excitation operator $T$ 
\begin{align}
    T &= \sum_m^N T_{m} \ , \notag \\
    T_{m} &= \frac{1}{(m!)^{2}}\sum_{ijab\ldots} \sum^{'}_{\substack{\kc_{a}\kc_{b}\\\kc_{i}\kc_{j}\ldots}}
    t_{i\kc_{i}j\kc_{j}\ldots}^{a\kc_{a}b\kc_{b}\ldots}a_{a\kc_{a}}^{\dagger}a_{b\kc_{b}}^{\dagger}\ldots a_{j\kc_{j}}a_{i\kc_{i}}\label{eq:ex_op}
\end{align}
generates all possible excitations from the occupied to virtual periodic molecular orbitals that are labeled using $i,j,k,l,\ldots$ and $a,b,c,d,\ldots$ indices, respectively.
The excitation amplitudes ($ t_{i\kc_{i}\kc_{j}\ldots}^{a\kc_{a}\kc_{b}\ldots}$) conserve crystal momenta, $\kc_{a} + \kc_{b} + \ldots - \kc_{i} - \kc_{j} - \ldots = \textbf{K}$, where $\textbf{K}$ is a reciprocal lattice vector.

Using the effective Hamiltonian $\tilde{H}$ and operator manifolds $h_{\pm\mu}^{\dagger}(\kc)$, expressions for the $n$th-order contributions to the EA/IP-ADC matrices can be written as:
\begin{align}
    M_{+\mu\nu}^{(n)}(\kc) &= \sum_{klm}^{k+l+m=n} \langle \Phi|[h_{+\mu}^{(k)}(\kc),[\tilde{H}^{(l)},h_{+\nu}^{(m)\dagger}(\kc)]]_{+}|\Phi\rangle \label{eq:M_ea} \ , \\
 M_{-\mu\nu}^{(n)}(\kc)&= \sum_{klm}^{k+l+m=n} \langle \Phi|[h_{-\mu}^{(k)\dagger}(\kc),[\tilde{H}^{(l)},h_{-\nu}^{(m)}(\kc)]]_{+}|\Phi\rangle\label{eq:M_ip}  \ , \\
T_{+p\nu}^{(n)}(\kc) &= \sum_{kl}^{k+l=n} \langle \Phi|[\tilde{a}_{p\kc}^{(k)},h_{+\nu}^{(l)\dagger}(\kc)]_{+}|\Phi\rangle \label{eq:T_ea} \ , \\
T_{-p\nu}^{(n)}(\kc) &= \sum_{kl}^{k+l=n} \langle \Phi|[\tilde{a}_{p\kc}^{(k)},h_{-\nu}^{(l)}(\kc)]_{+}|\Phi\rangle\label{eq:T_ip} \ , 
\end{align}
where $[\ldots]$ and $[\ldots]_{+}$ denote a commutator and an anti-commutator, respectively. 
The low-order EA-ADC($n$) approximations ($n\le3$) require two classes of electron-attachment operator manifolds, $h_{+\mu}^{(0)\dagger}(\kc) = a_{a\kc}^{\dagger}$ and ${h}_{+\mu}^{(1)\dagger}(\kc) =a_{b\kc_{b}}^{\dagger}a_{a\kc_{a}}^{\dagger}a_{i\kc_{i}}$ with $\kc_{a} + \kc_{b} - \kc_{i} = \textbf{k}$.
Similarly, the IP-ADC($n$) ($n\le3$) methods employ two ionization operator manifolds $h_{-\mu}^{(0)\dagger}(\kc) = a_{i\kc}$ and ${h}_{-\mu}^{(1)\dagger}(\kc) =a_{a\kc_{a}}^{\dagger}a_{j\kc_{j}}a_{i\kc_{i}}$ with \kc$_{a}$ $-$ \kc$_{i}$ $-$ \kc$_{j} = \textbf{k}$. 

Evaluating \cref{eq:M_ea,eq:T_ea,eq:M_ip,eq:T_ip} also requires the expressions for $k$th-order contributions to the effective Hamiltonian $\tilde{{H}}$ and observable $\tilde{a}_{p\kc}=e^{-(T-T^{\dagger})}a_{p\kc}e^{(T-T^{\dagger})}$, which are obtained from a perturbative analysis of the Baker--Campbell--Hausdorff expansions
\begin{align}
    \tilde{H} &= H^{(0)} + V + [H^{(0)},T^{(1)} - T^{(1){\dagger}}] \notag \\
    &+ [H^{(0)},T^{(2)} - T^{(2){\dagger}}]\notag   \\
    &+\frac{1}{2!}[2V + [H^{(0)},T^{(1)} - T^{(1){\dagger}}],T^{(1)}-T^{(1){\dagger}}]+... \ , \label{eq:H_eff}\\ 
    \tilde{a}_{p\kc} &= a_{p\kc} + [a_{p\kc},T^{(1)}-T^{(1){\dagger}}]+ [a_{p\kc},T^{(2)}-T^{(2){\dagger}}]\notag\\
    &+\frac{1}{2!}[[a_{p\kc},T^{(1)}-T^{(1){\dagger}}],T^{(1)}-T^{(1){\dagger}}]+... \ . \label{eq:ap_eff}
\end{align}
The excitation operators $T^{(k)}$ in \cref{eq:H_eff,eq:ap_eff} depend on the amplitudes $t_{i\kc_{i}j\kc_{j}\ldots}^{a\kc_{a}b\kc_{b}\ldots{(k)}}$ that can be computed at any order $k$ by solving a system of projected amplitude equations:
\begin{align}
    \langle\Phi| a_{i\kc_{i}}^{\dagger}a_{a\kc_{a}}\tilde{H}^{(k)} |\Phi\rangle &= 0,\label{eq:T_amp_s} \\
    \langle\Phi| a_{i\kc_{i}}^{\dagger}a_{j\kc_{j}}^{\dagger}a_{b\kc_{b}}a_{a\kc_{a}}\tilde{H}^{(k)} |\Phi\rangle &= 0, \ \ldots \label{eq:T_amp_d}
\end{align}
These $k$th-order excitation amplitudes are equivalent to the amplitudes of $k$th-order wavefunction in the periodic M\o ller--Plesset perturbation theory. \cite{Moller:1934kp618,Sun:1996p8553,Ayala:2001p9698} 
For example, the first-order double-excitation and second-order single-excitation amplitudes used in the EA/IP-ADC(2) methods have the form:
\begin{align}
 t_{i\kc_{i}j\kc_{j}}^{a\kc_{a}b\kc_{b}(1)} &= \frac{v^{a\kc_{a}b\kc_{b}}_{i\kc_{i}j\kc_{j}}}{\e{i\kc_{i}}+\e{j\kc_{j}}-\e{a\kc_{a}}-\e{b\kc_{b}}} \ , \label{eq:t2_1}
\end{align}
\begin{align}
({\e{i\kc_{i}} -\e{a\kc_{a}}}) t_{i\kc_{i}}^{a\kc_{a}(2)}  &=\frac{1}{2}\sum\limits_{kcd}\sum\limits^{'}_{\substack{\kc_{k}\kc_{c} \\ \kc_{d}}}v^{a\kc_{a}k\kc_{k}}_{c\kc_{c}d\kc_{d}}t_{i\kc_{i}k\kc_{k}}^{c\kc_{c}d\kc_{d}(1)} \notag \\
&- \frac{1}{2}\sum\limits_{klc}\sum\limits^{'}_{\substack{\kc_{k}\kc_{l} \\ \kc_{c}}}v^{k\kc_{k}l\kc_{l}}_{i\kc_{i}c\kc_{c}}t_{k\kc_{k}l\kc_{l}}^{a\kc_{a}c\kc_{c}(1)} \ .
 \label{eq:t1_2}
\end{align}
Equations for other low-order amplitudes are reported in the Supporting Information. 

\begin{figure*}[t!]
	\includegraphics[width=5in]{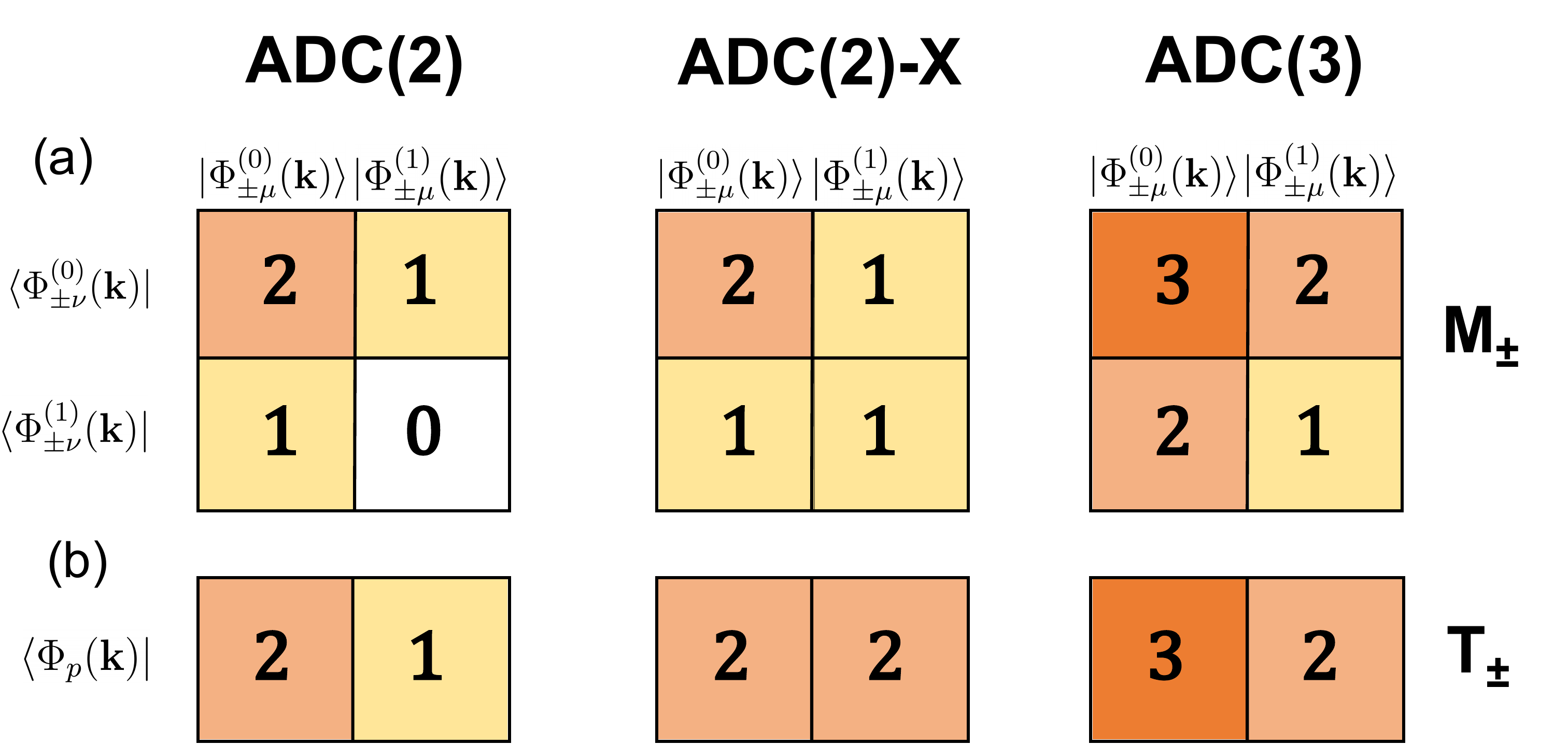}
	\captionsetup{justification=raggedright,singlelinecheck=false,font=footnotesize}
	\caption{Perturbative structure of the (a) effective Hamiltonian $\mathbf{M_\pm}$ and (b) effective transition moments $\mathbf{T_\pm}$ matrices in periodic EA/IP-ADC(2), EA/IP-ADC(2)-X, and EA/IP-ADC(3).}
	\label{fig:adc_matrices}
\end{figure*}

In this work, we consider three periodic EA/IP-ADC($n$) ($n\le$ 3) approximations, namely: EA/IP-ADC(2), EA/IP-ADC(2)-X, and EA/IP-ADC(3). 
The perturbative structure of their effective Hamiltonian and transition moments matrices is shown in \cref{fig:adc_matrices}. 
The EA/IP-ADC(2) methods include all contributions to $ \textbf{M}_{\pm}(\kc)$ and $ \textbf{T}_{\pm}(\kc)$ strictly up to the second order in perturbation theory. 
The EA/IP-ADC(2)-X approximation is an extension to ADC(2), where the diagonal block evaluated with respect to the first-order excitations $h_{\pm\mu}^{\dagger(1)}(\kc)$ includes the third-order terms. 
While both EA/IP-ADC(2) and EA/IP-ADC(2)-X employ the MP2 description of electron correlation in the ground and singly-excited ($\ket{\Phi_{\pm\mu}^{(0)}(\kc)} = h_{\pm\mu}^{(0)\dagger}(\kc) \ket{\Phi}$) electronic states, the extended approximation provides a higher-level description of doubly-excited (shake-up) states ($\ket{\Phi_{\pm\mu}^{(1)}(\kc)} = h_{\pm\mu}^{(1)\dagger}(\kc) \ket{\Phi}$) and orbital relaxation effects than EA/IP-ADC(2). 
The EA/IP-ADC(3) methods incorporate all third-order contributions to 1-GF, including the MP3 description of ground-state electron correlation and third-order treatment of singly-excited states ($\ket{\Phi_{\pm\mu}^{(0)}(\kc)}$).
We present the $k$-adapted spin-orbital working equations for all periodic EA/IP-ADC($n$) ($n\le$ 3) matrix elements and amplitudes in the Supporting Information. 

\section{Implementation}
\label{sec:implementation}

We implemented the $k$-adapted periodic EA/IP-ADC methods as part of the periodic boundary conditions (PBC) module in the \textsc{PySCF} program.\cite{Sun:2020p024109}
This new implementation of EA/IP-ADC for crystalline systems is an extension of our efficient EA/IP-ADC program for molecules,\cite{Banerjee:2021p074105} with two major modifications: (i) adding the support of complex-valued basis sets such as the periodic molecular orbital (MO) basis used in this work (\cref{eq:crys_MO}) and (ii) incorporating crystalline translational symmetry by taking advantage of crystal momentum conservation.
The $k$-point EA/IP-ADC implementation was verified against an inefficient $\Gamma$-point implementation, which does not support translational symmetry, but can be used to perform the periodic EA/IP-ADC calculations at the $\Gamma$-point of the Brillouin zone ($\textbf{k}=(0,0,0)$) using a real-valued Gaussian basis set. 

The periodic EA/IP-ADC calculations start by defining a unit cell of a crystal, an atomic orbital (AO) basis set (with pseudopotential, if necessary), and a mesh of $k$-points.
Next, a periodic restricted Hartree--Fock calculation is performed to compute the crystalline MO's ($\psi_{p\kc}(\textbf{r}) $, \cref{eq:crys_MO}) and the orbital energies ($\epsilon_{p}$, \cref{eq:H_0}), followed by the transformation of two-electron integrals from AO to periodic MO basis ($v_{p\kc_{p}q\kc_{q}}^{r\kc_{r}s\kc_{s}}$).
To reduce the cost of two-electron integral transformation and storage, our EA/IP-ADC implementation supports Gaussian density fitting,\cite{Whitten:1973p4496,Dunlap:1979p3396,Vahtras:1993p514,Feyereisen:1993p359} which factorizes the four-index integral tensors into the product of two- and three-index tensors that can be stored and transformed more efficiently.
In our previous work, we have demonstrated that the density-fitting approximation has a negligible effect on the EA/IP-ADC excitation energies (with errors of $\sim$ $10^{-3}$ eV or less).\cite{Banerjee:2021p074105}

The transformed two-electron integrals and orbital energies are used to compute the amplitudes of EA/IP-ADC($n$)/ADC($n$)-X effective Hamiltonian $\tilde{H}$ (\cref{sec:theory:periodic_adc_theory}), which are obtained by performing the periodic $n$th-order M\o ller--Plesset perturbation theory calculation (MP$n$, $n$ = 2, 3) for the ground electronic state.
To reduce the computational cost of calculating the amplitudes, we neglect the contributions from second-order singles ($t_{i\kc_{i}}^{a\kc_{a}(2)}$) in the EA/IP-ADC(2) equations, second-order singles and doubles ($t_{i\kc_{i}j\kc_{j}}^{a\kc_{a}b\kc_{b}(2)}$) in EA/IP-ADC(2)-X, and third-order singles ($t_{i\kc_{i}}^{a\kc_{a}(3)}$) in EA/IP-ADC(3). 
These approximations do not affect the computed charged excitation energies and have a minor effect on the calculated spectroscopic factors.
We also note that our molecular and periodic EA/IP-ADC implementations neglect the contribution of triple excitations that formally enter the expressions for $\textbf{T}_{\pm}$ at the third order in perturbation theory.\cite{Hodecker:2022p074104}

Once the amplitudes of effective Hamiltonian are computed and stored on disk, our periodic EA/IP-ADC implementation calculates the band energies for each $k$-point by iteratively solving the eigenvalue problem in \cref{eq:eig} using the multiroot Davidson algorithm\cite{Davidson:1975p87,Liu:1978p49}.
If requested, the computed eigenvectors are used to calculate the momentum-resolved spectroscopic factors (\cref{eq:spec_factors}) and crystalline density of states (\cref{eq:dos}).

Taking advantage of the crystalline translational symmetry and density fitting, the computational cost of our $k$-adapted periodic EA/IP-ADC program is dominated by the calculation of effective Hamiltonian amplitudes, which scales as $\mathcal{O}(N_{k}^3N_{aux}N_O^2N_V^2)$ for EA/IP-ADC(2) or EA/IP-ADC(2)-X and $\mathcal{O}(N_{k}^4N_O^2N_V^4)$ for EA/IP-ADC(3), where $N_{k}$ is the total number of $k$-points, while $N_O$, $N_V$, and $N_{aux}$ denote the number of occupied, virtual, and auxillary basis functions per unit cell, respectively. 
By precomputing and reusing efficient tensor intermediates, calculating charged excitation energies and spectroscopic factors has a computational scaling of $\mathcal{O}(N_{k}^3N_O^3N_V^2)$ and $\mathcal{O}(N_{k}^3N_O^2N_V^3)$ for all IP- and EA-ADC methods developed in this work, respectively.
This offers a  $\mathcal{O}(N_{k}^2)$ savings compared to the $\Gamma$-point EA/IP-ADC implementations.  

\section{Computational Details}
\label{sec:computational_details}

The $k$-point periodic MP2, MP3, and EA/IP-ADC methods were implemented in the \textsc{PySCF}\cite{Sun:2020p024109} software package as described in \cref{sec:implementation}.
All calculations employed the Gaussian density fitting with the \textsc{PySCF}-autogenerated even-tempered auxiliary basis sets and a uniform Monkhorst--Pack mesh\cite{Monkhorst:1976p5188} to sample the first Brillouin zone.

In \cref{sec:results:equation_of_state}, we present the equation-of-states and lattice parameters of diamond (C) and silicon (Si) computed using our $k$-point MP2 and MP3 programs and compare to the reference results from periodic coupled cluster theory with single and double excitations (CCSD)\cite{Hirata:2004p2581,McClain:2017p1209} implemented in \textsc{PySCF}.
C and Si share the same crystal structure, with two atoms per primitive unit cell.
The equation-of-state calculations were performed using the gth-cc-pVTZ basis set and the gth-hf-rev pseudopotential\cite{Hong-Zhou:2022p1595} with a 3 $\times$ 3 $\times$ 3 $k$-point mesh. 
Lattice constants at 0 K were obtained by fitting the results using a third-order Birch--Murnaghan curve.\cite{Birch:1947p809}

In \cref{sec:results:band_structures}, we report the band structures of C and Si computed using periodic EA/IP-ADC methods and compare to the band structures from density functional theory (PBE0 functional)\cite{Perdew:1996p9982,Adamo:1999p6158} and equation-of-motion coupled cluster theory with single and double excitations (EOM-CCSD).\cite{McClain:2017p1209}
These calculations were performed using the 3 $\times$  3 $\times$ 3 $k$-point mesh, employed the gth-DZVP basis and gth-pade pseudopotential,\cite{Goedecker:1996p1703,Vandevondele:2007p114105} and used the experimental lattice parameters obtained at 300 K: $a_{\mathrm{C}}$ = 3.567 $\angstrom$ and $a_{\mathrm{Si}}$ = 5.431 $\angstrom$.

Finally, in \cref{sec:results:band_gaps}, we benchmark the accuracy of periodic EA/IP-ADC methods for calculating the thermodynamic-limit band gaps of seven crystalline materials: diamond (C), silicon (Si), silicon carbide (SiC), lithium fluoride (LiF), magnesium oxide (MgO), neon (Ne), and argon (Ar). 
SiC has a diamond-like crystal structure, while LiF and MgO crystallize in a rocksalt cubic structure. 
The solid Ne and Ar share a face-centered cubic crystal structure. 
Band gaps were computed using the experimental (300 K) lattice parameters, specifically: $a_{\mathrm{C}}$ = 3.567 $\angstrom$, $a_{\mathrm{Si}}$ = 5.431 $\angstrom$, $a_{\mathrm{SiC}}$ = 4.350 $\angstrom$, $a_{\mathrm{LiF}}$ = 4.035 $\angstrom$, $a_{\mathrm{MgO}}$ = 4.213 $\angstrom$, $a_{\mathrm{Ar}}$ = 5.256 $\angstrom$, and $a_{\mathrm{Ne}}$ = 4.429 $\angstrom$.\cite{Madelung:2004semiconductors}
For all materials, bands gaps were calculated as the difference between conduction band minimum and valence band maximum in the band structure.
For C, Si, and SiC, the computed band gaps are indirect, i.e.\@ involve a change in the crystal momentum, while the remaining materials have a direct band gap.
The calculations of band gaps were performed using the gth-cc-pV$X$Z basis sets with the gth-hf-rev pseudopotential ($X$ = D, T) for all materials except for Ar and Ne, which used a gth-aug-cc-pV$X$Z basis set ($X$ = D, T)\cite{Hong-Zhou:2022p1595} with the same pseudopotential.

The EA/IP-ADC band gaps were extrapolated to the thermodynamic limit assuming that the finite-size error decays as $\mathcal{O}(N_{k}^{-1/3})$ with the increasing number of $k$-points $N_{k}$.\cite{Aissing:1993p81,Sundararaman:2013p165122,Broqvist:2009p085114,Shepherd:2013p226401,McClain:2017p1209}
Using EA/IP-ADC(2) and the gth-cc-pVDZ basis set, we explored two schemes: i) a two-point extrapolation with $N_{k}$ = 3 $\times$ 3 $\times$ 3 and 4 $\times$ 4 $\times$ 4, and ii) a three-point extrapolation with $N_{k}$ = 3 $\times$ 3 $\times$ 3, 4 $\times$ 4 $\times$ 4, and 5 $\times$ 5 $\times$ 5.
The results of the extrapolation schemes differed by $\le$ 0.1 eV for all materials except SiC, where a difference of 0.27 eV was obtained (see Table S1 in the Supplementary Information for details).
Since the two-point $3^3-4^3$ extrapolation provided satisfactory results, we adopted this scheme for the extrapolation of band gaps computed using EA/IP-ADC(2)/gth-cc-pVTZ, EA/IP-ADC(2)-X/gth-cc-pV$X$Z ($X$ = D, T), and EA/IP-ADC(3)/gth-cc-pVDZ. 
The EA/IP-ADC(3)/gth-cc-pVTZ thermodynamic-limit band gaps were estimated by making an additivity assumption 
\begin{align}
\label{eq:adc3_tz_gap}
E_{g,\infty}^{ADC(3)}  (TZ) &=  E_{g,\infty}^{ADC(3)} (DZ) \\ \notag
&+
 E_{g,3\times 3 \times 3}^{ADC(3)} (TZ)  -  E_{g,3\times 3 \times 3}^{ADC(3)}  (DZ) \ ,
\end{align}
where $ E_{g,\infty}^{ADC(3)} (DZ)$ is the band gap extrapolated at the EA/IP-ADC(3)/gth-cc-pVDZ level of theory, while $ E_{g,3\times 3 \times 3}^{ADC(3)} (DZ)$ and $ E_{g,3\times 3 \times 3}^{ADC(3)} (TZ)$ are the EA/IP-ADC(3)/gth-cc-pV$X$Z ($X$ = D, T) band gaps calculated using the 3 $\times$ 3 $\times$ 3 $k$-point mesh.

\section{Results}
\label{sec:results}

\subsection{Ground-state properties}
\label{sec:results:equation_of_state}

\begin{figure*}[t!]
	\subfigure[]{\includegraphics[width=2.8in]{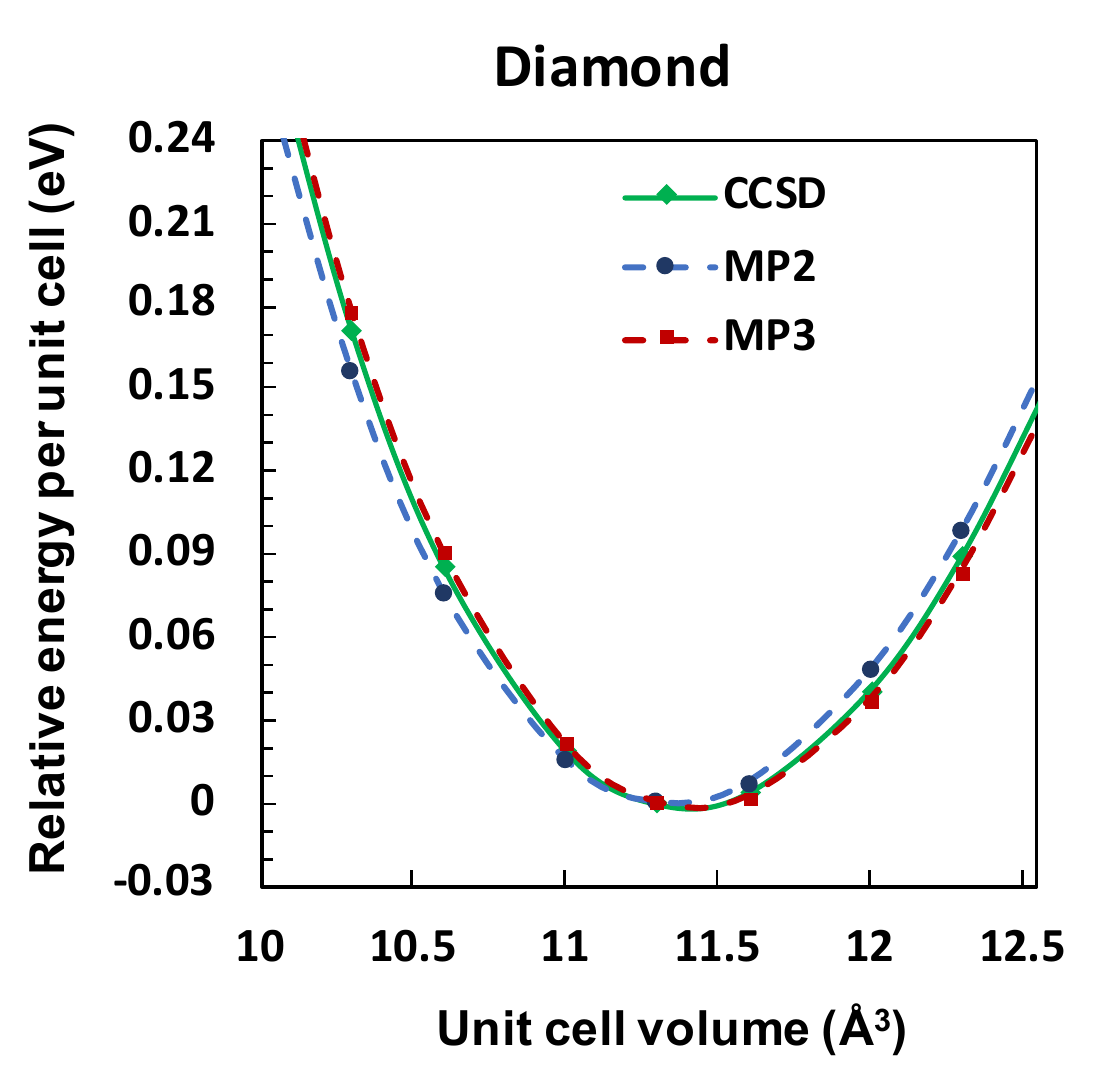}\label{fig:diamond_eos}}  
	\subfigure[]{\includegraphics[width=2.8in]{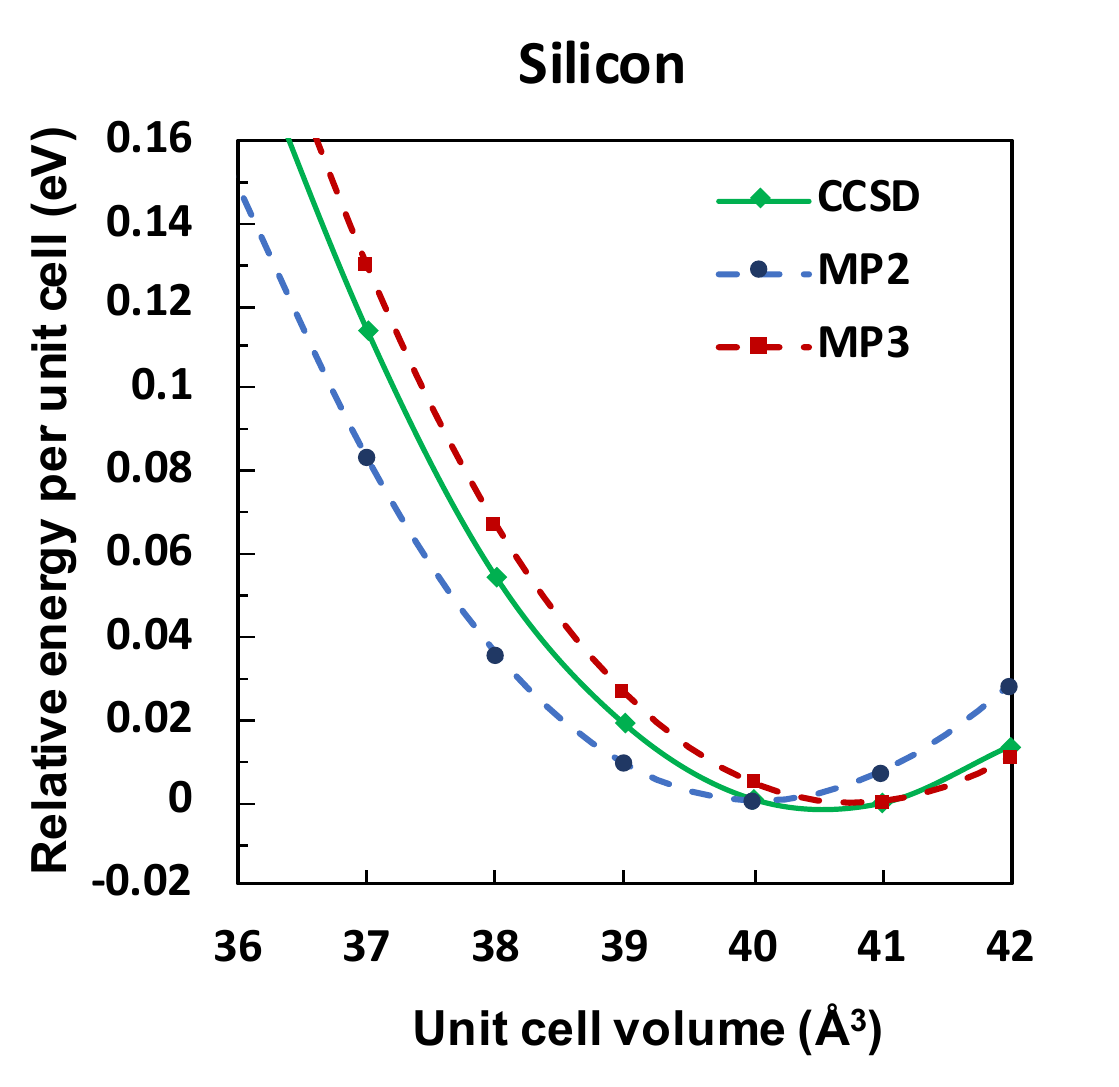}\label{fig:si_eos}}  
	\captionsetup{justification=raggedright,singlelinecheck=false,font=footnotesize}
	\caption{Equation-of-state for diamond (a) and silicon (b) computed using MP2, MP3, and CCSD with the gth-cc-pVTZ basis set, gth-hf-rev pseudopotential, and 3 $\times$ 3 $\times$ 3 $k$-point mesh. 
	}
	\label{fig:equation_of_state}
\end{figure*}

First, we benchmark the accuracy of periodic MP2 and MP3 methods for simulating the equation-of-state (EOS) and lattice constants of crystalline diamond (C) and silicon (Si). 
Although periodic MP2 has been previously used to compute the ground-state properties of three-dimensional crystalline materials,\cite{Sun:1996p8553,Ayala:2001p9698,Marsman:2009p184103,Gruneis:2010p074107,Gruneis:2011p2780,Usvyat:2013p194101,Shepherd:2012p035111,McClain:2017p1209,Lange:2021p081101} the applications of periodic MP3 have been limited to one- and two-dimensional systems.\cite{Suhai:1982p3506,Suhai:1993p131,Suhai:1994p14791,Hirata:2004p2581,Keceli:2010p115107}
\cref{fig:equation_of_state} shows the EOS of C and Si calculated by plotting the ground-state energy per unit cell, relative to its equilibrium value, as a function of the unit cell volume. 
For C, the EOS from MP2 and MP3 are very close to the EOS computed using coupled cluster theory with single and double excitations (CCSD), which we use here as a reference. 
The EOS of Si represents a more challenging test for finite-order perturbation theory, with MP2 and MP3 showing noticeable deviations from CCSD. 
For both materials, the third-order correlation energy correction in MP3 significantly improves the description of crystalline ground-state energies relative to MP2, moving the computed EOS closer to that of CCSD. 

\begin{table}[t!]
	\captionsetup{justification=raggedright,singlelinecheck=false,font=footnotesize}
	\caption{Lattice parameters ($a$, $\angstrom$) of diamond (C) and silicon (Si) at 0 K computed from \cref{fig:equation_of_state}. 
		Also reported are the experimental lattice parameters corrected for the zero-point vibrational effects.\cite{Schimka:2011p024116}}
	\label{tab:lattice_parameters}
	\footnotesize
	\begin{tabular}{C{2cm}C{2cm}C{2cm}}
		\hline
		\hline
		\multicolumn{1}{c}{Method} &\multicolumn{1}{c}{$a$(C)} &\multicolumn{1}{c}{$a$(Si)}  \\
		\hline
		MP2                         &3.567 		&5.430		 \\                    	
		MP3                          &3.574		&5.463		 \\
		CCSD   						&3.572 		&5.455		 \\
		Experiment 				&3.553         &5.421 	\\
		\hline
		\hline
	\end{tabular}
\end{table}

The EOS plots provide the equilibrium volumes, which can be used to calculate the equilibrium lattice constants at 0 K. 
\cref{tab:lattice_parameters} reports the equilibrium lattice parameters computed using MP2, MP3, and CCSD.
For comparison, we also report the experimental lattice constants corrected for the zero-point vibrational effects.\cite{{Schimka:2011p024116}}
For C, the best agreement with CCSD is demonstrated by MP3, which overestimates the lattice parameter by only 0.002 $\angstrom$, while MP2 underestimates this equilibrium constant by 0.005 $\angstrom$.
The differences in results are more prominent for Si, where the MP2 error in lattice constant is 0.025 $\angstrom$. 
The MP3 method reduces the error in Si lattice parameter to 0.008 $\angstrom$.

Overall, our results demonstrate that periodic MP3 is significantly more accurate than MP2 and is competitive with CCSD in the description of crystalline ground-state properties.
This, together with a lower computational cost, makes MP3 an attractive alternative to CCSD for simulating the ground-state properties of weakly-correlated crystalline systems.

\subsection{Band structures}
\label{sec:results:band_structures}

Turning to excited-state properties, in this section we use the $k$-point EA/IP-ADC methods to simulate the crystalline band structures of C and Si and compare these results to the band structures from periodic Hartree--Fock theory (HF), density functional theory (PBE0 functional), and equation-of-motion CCSD (EOM-CCSD) reported in Ref.\@ \citenum{McClain:2017p1209}. 
Band structures provide important information about the electronic structure of a material by plotting the energies of charged electronic states as a function of crystal momentum $k$.
For each material, we compute the three highest valence (IP) and four lowest conduction (EA) bands by performing independent IP- and EA-ADC calculations for a fixed number of $k$-points in the first Brillouin zone.
All band structure calculations were performed using the 3 $\times$ 3 $\times$ 3 $k$-point mesh with the gth-DZVP basis and gth-pade pseudopotential.

\begin{figure*}[t!]
	\subfigure[]{\includegraphics[width=2.8in]{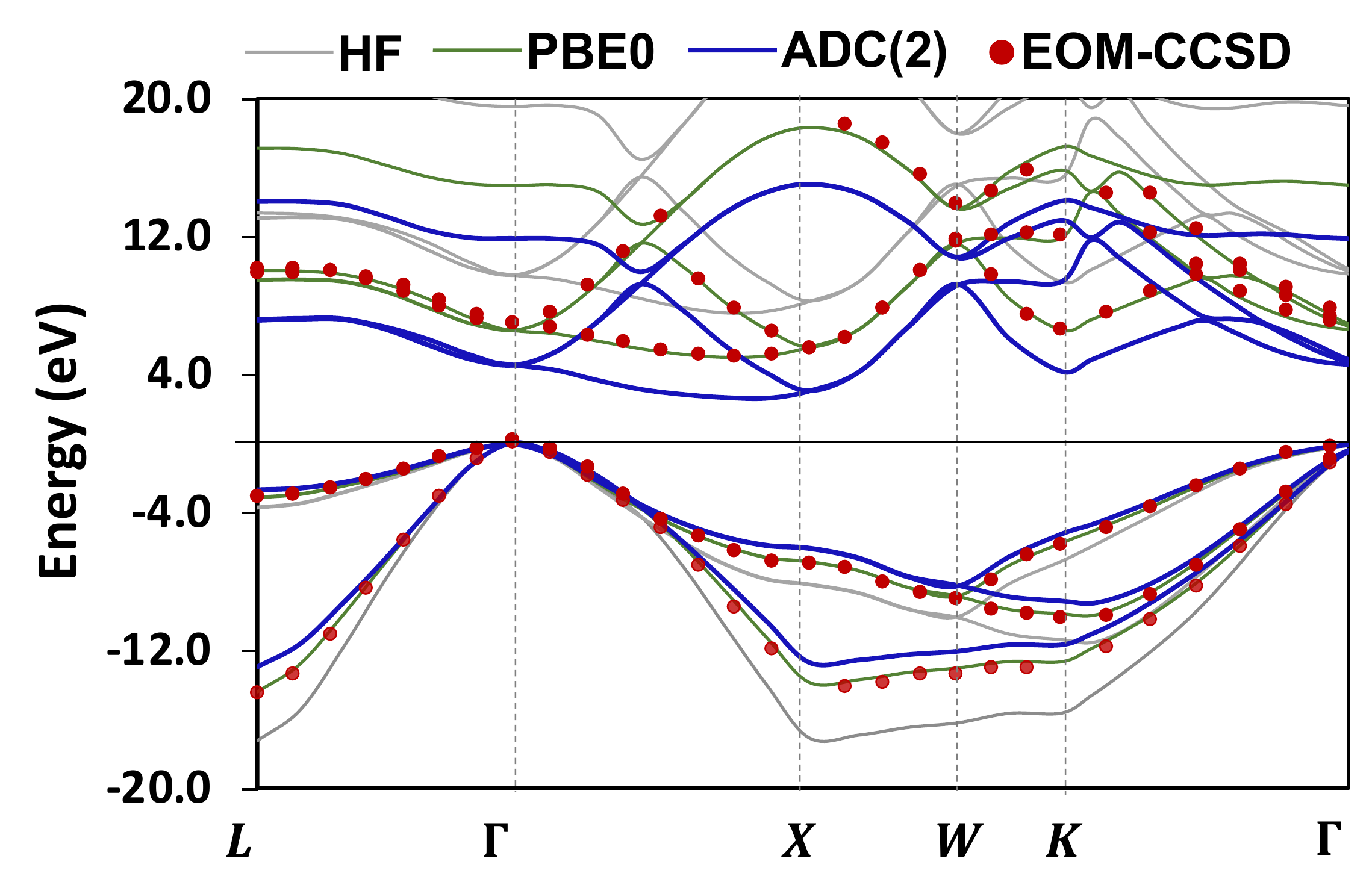}\label{fig:c_eos_2}}  
	\subfigure[]{\includegraphics[width=2.8in]{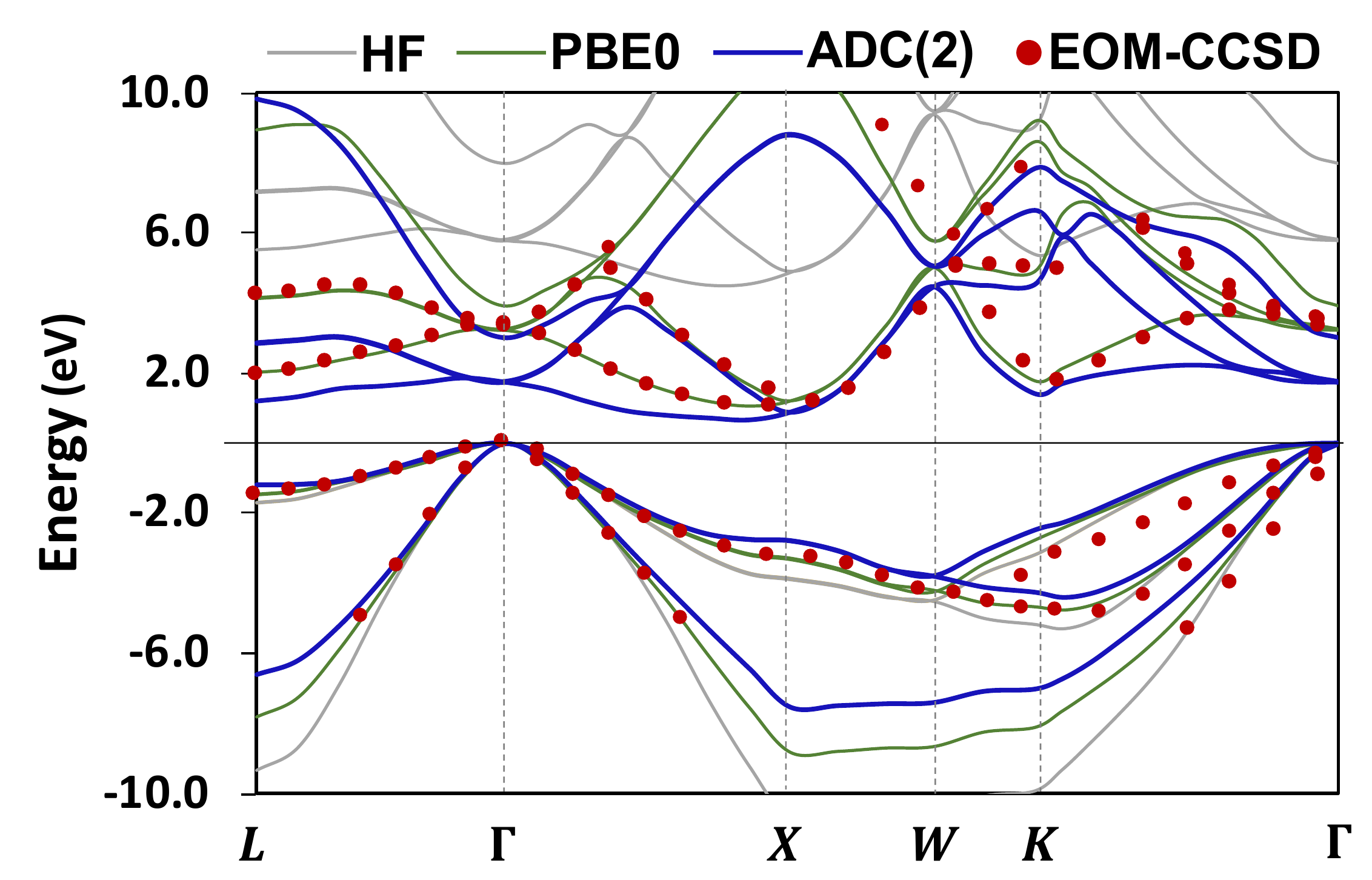}\label{fig:si_eos_2}}  
	\medskip
	\subfigure[]{\includegraphics[width=2.8in]{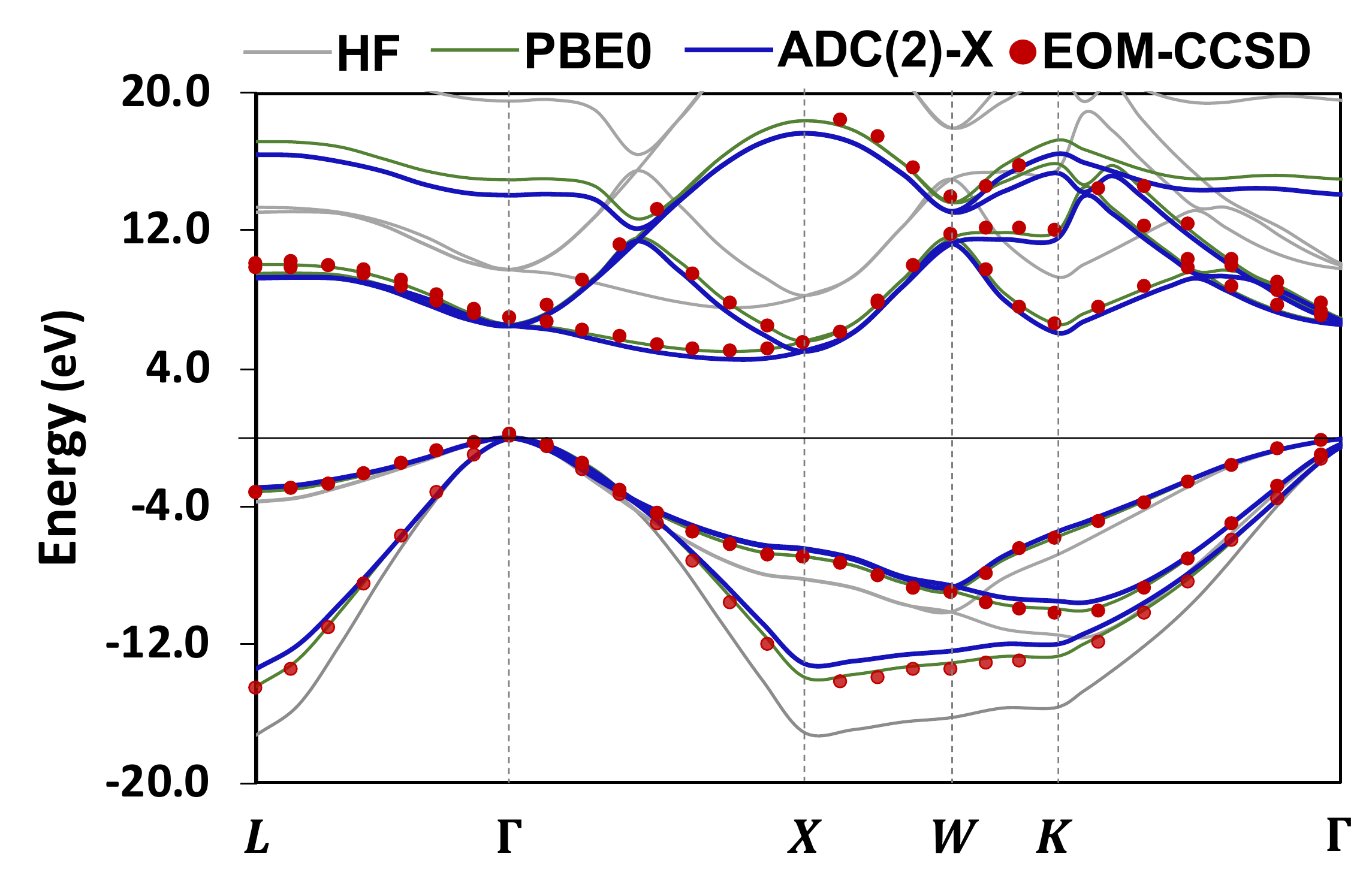}\label{fig:c_eos_2x}}  
	\subfigure[]{\includegraphics[width=2.8in]{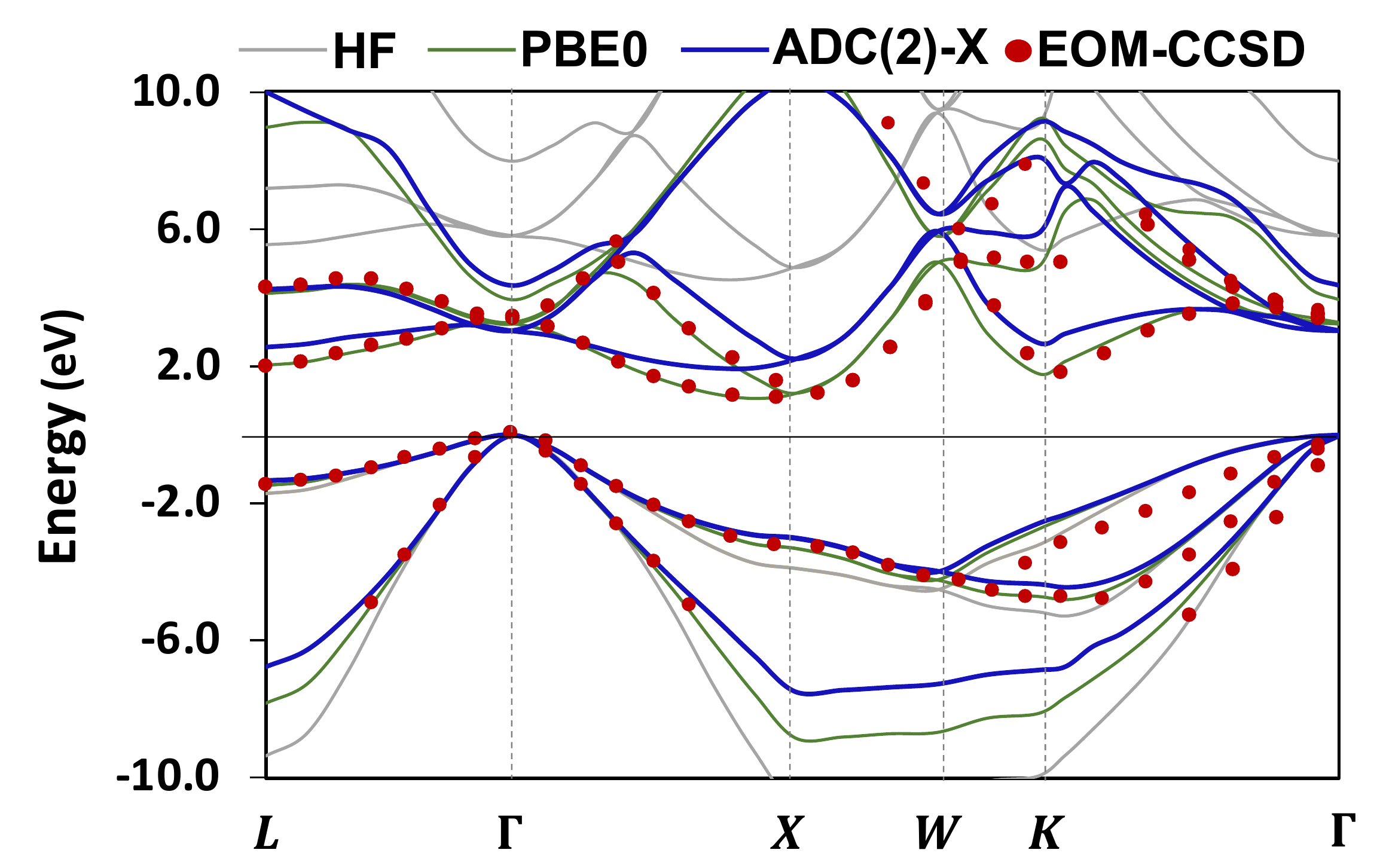}\label{fig:si_eos_2x}}  
	\medskip
	\subfigure[]{\includegraphics[width=2.8in]{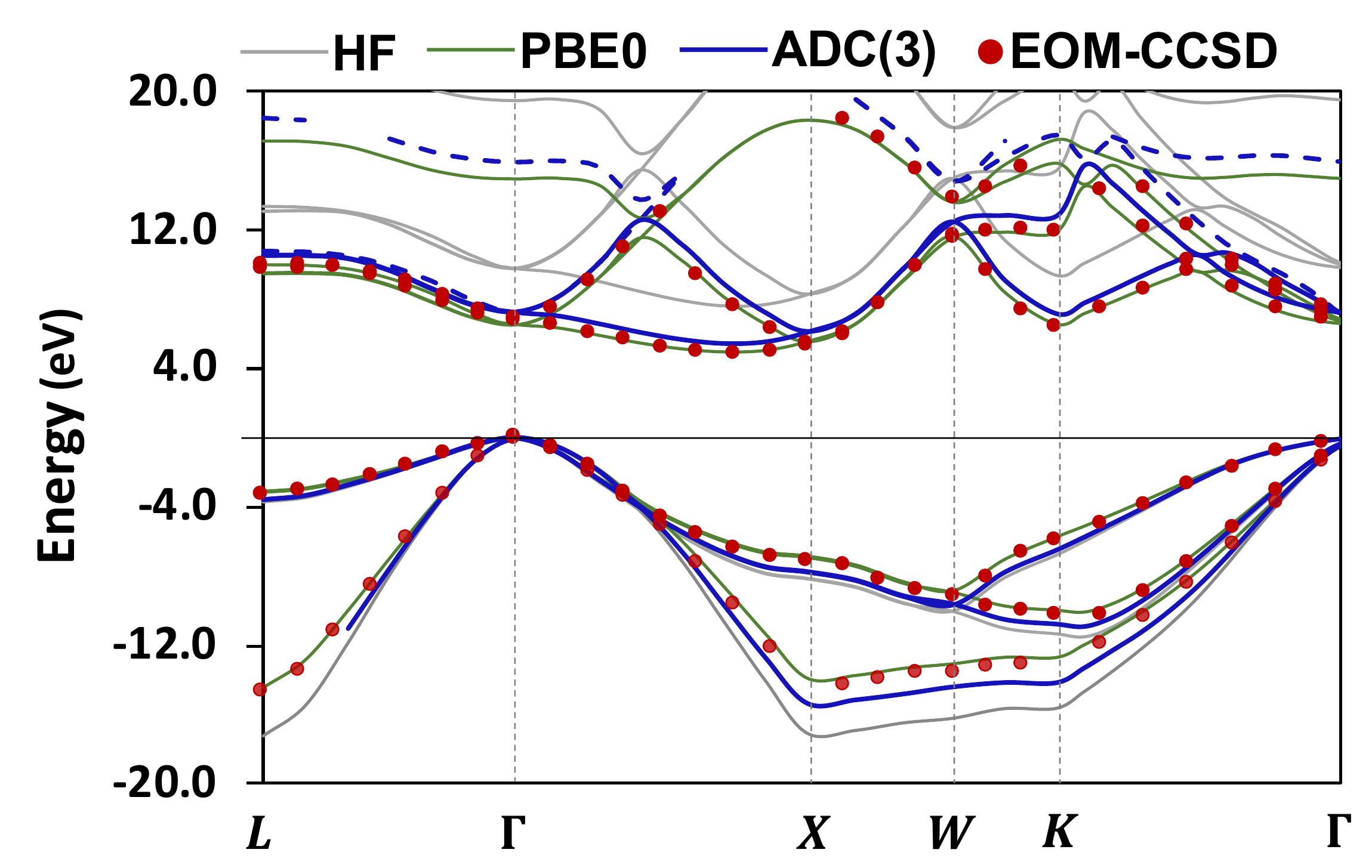}\label{fig:c_eos_3}}  
	\subfigure[]{\includegraphics[width=2.8in]{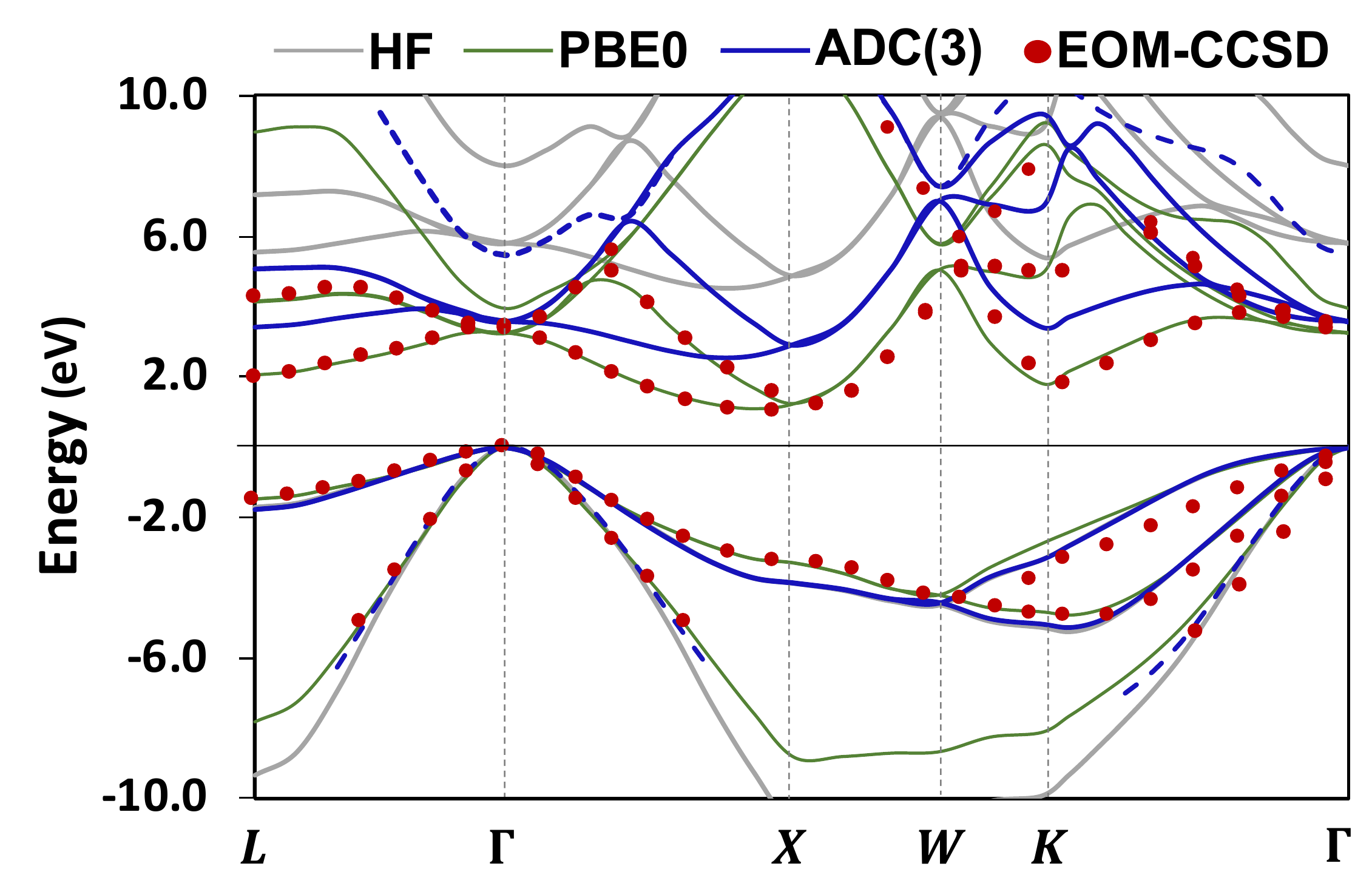}\label{fig:si_eos_3}}
	\captionsetup{justification=raggedright,singlelinecheck=false,font=footnotesize}
	\caption{Band structures of diamond (a, c, e) and silicon (b, d, f) computed using the periodic EA/IP-ADC methods with the gth-DZVP basis set, gth-pade pseudopotential, and $3\times3\times3$ sampling of the Brillouin zone. Results are compared to the band structures calculated using the Hartree--Fock theory (HF), density functional theory (PBE0), and equation-of-motion coupled cluster theory (EOM-CCSD).\cite{McClain:2017p1209} Due to convergence problems, some EA/IP-ADC(3) bands are incomplete as indicated by the dashed lines. }
	\label{fig:band_structure}
\end{figure*}

\cref{fig:c_eos_2,fig:c_eos_2x,fig:c_eos_3} show the band structures of C computed using EA/IP-ADC(2), EA/IP-ADC(2)-X, and EA/IP-ADC(3), respectively.
Employing a finite basis set and $k$-point mesh, all ADC methods produce qualitatively similar band structures but differ in the prediction of energy gap between the conduction and valence bands.
The EA/IP-ADC(2) methods correctly reproduce all features of the band structure from EOM-CCSD but consistently underestimate the conduction band energies in the sampled Brillouin zone by $\sim$ 2.3 eV.
Incorporating the third-order correlation effects in EA/IP-ADC(2)-X and EA/IP-ADC(3) opens up the band gap, reducing the errors in conduction band energies to $\sim$ $-$0.4 and 0.7 eV, respectively.
All three methods show a significant improvement over the HF theory, which exhibits a non-uniform performance in predicting the band energies in the Brillouin zone.

The band structures of Si computed using EA/IP-ADC are shown in \cref{fig:si_eos_2,fig:si_eos_2x,fig:si_eos_3}. 
As for the ground-state properties (\cref{sec:results:equation_of_state}), simulating charged excited states of Si is a more challenging test for finite-order perturbation theories such as ADC, due to a much smaller fundamental band gap in this material compared to C.
Although the EA/IP-ADC methods reproduce all qualitative features of the EOM-CCSD band structure, their performance is not uniform across the Brillouin zone.
The EA/IP-ADC(2) methods significantly underestimate the conduction band energies near the $\Gamma$- and $L$-points but show much closer agreement to EOM-CCSD along the $X-W$ region and near the $K$-point. 
In contrast, EA/IP-ADC(2)-X and EA/IP-ADC(3) exhibit a good performance near the $\Gamma$-point, but overestimate the conduction band energies in other regions, with the exception of $L-\Gamma$ path for EA/IP-ADC(2)-X.
This apparent lack of convergence in the ADC perturbation series underscores the importance of high-order electron correlation effects for small-gap semiconductors like Si.

\begin{table*}[t!]
	\begin{threeparttable}
		\captionsetup{justification=raggedright,singlelinecheck=false,font=footnotesize}
		\caption{Minimum fundamental band gaps (eV) calculated using EA/IP-ADC(2), EA/IP-ADC(2)-X, and EA/IP-ADC(3) with the gth-cc-pV$X$Z ($X$ = D, T) basis sets and gth-hf-rev pseudopotential. 
			For Ne and Ar, the gth-aug-cc-pV$X$Z ($X$ = D, T) basis sets were used.
			Band gaps were extrapolated to the thermodynamic limit as described in \cref{sec:computational_details}.
		}
		\label{tab:band_gaps}
		\footnotesize
		\setstretch{1}
		\begin{tabular}{L{1.2cm}C{1.2cm}C{1.2cm}C{1.2cm}C{1.2cm}C{1.2cm}C{1.2cm}C{1.2cm}C{1.5cm}C{1.5cm}}
			\hline
			\hline
			\multicolumn{1}{c}{Material} &\multicolumn{2}{c}{ADC(2)} &\multicolumn{2}{c}{ADC(2)-X} &\multicolumn{2}{c}{ADC(3)} &\multicolumn{1}{c}{MP2\tnote{b}} &\multicolumn{1}{c}{Experiment\tnote{c}}\\
			&DZ &TZ &DZ &TZ &DZ &TZ\tnote{a}   \\
			\hline 
			Si		&$-$3.72	&$-$3.72	 &$-$0.13	&$-$0.19	&3.12	&3.18	&$-$2.13           &1.30 \\ 
			SiC		&$-$2.66	&$-$2.31	 &0.74	        &0.79	        &4.97	&5.05	&$-$1.21           &2.37 \\
			C	        &$-$0.14	&$-$0.21	 &3.76	        &3.68	        &8.05	&7.99	&1.57	           &5.81  \\
			MgO	        &6.39	        &5.99 &6.99	        &6.84	        &9.85	&9.82	&7.70	           &8.19  \\
			Ar	        &13.54	        &13.33	       &13.33	        &13.23	        &14.79	&14.64	&13.80	   	   &14.20 \\
			LiF	        &14.06	        &13.78	        &14.18	        &14.02        &16.54	&16.53	&14.88	           &15.09 \\
			Ne	        &20.53	        &20.46	        &20.76	        &20.64	        &22.37	&22.24	&21.00	            &21.70\\
			\hline
			\mae	        &2.95      &3.05	        &1.29	        &1.38	        &1.54			&1.58				&1.86	            &\\
			\std	        &2.27       &2.13	        &0.44		        &0.40        &0.75			&0.82				&1.79	            &\\
			\maxe	        &5.95		        &6.02	        &2.05		        &2.13		        &2.6			&2.68					&4.24	            &\\
			\hline
			\hline		
		\end{tabular}
		\begin{tablenotes}
			\item[a] The ADC(3)/gth-cc-pVTZ band gaps were estimated using \cref{eq:adc3_tz_gap}.
			\item[b] MP2 band gaps calculated using the all-electron cc-pVTZ basis set from Ref.\@ \citenum{Lange:2021p081101}. 
			\item[c] The experimental band gaps\cite{Madelung:2004semiconductors,Chiang:1989electronic,Schwentner:1975p528} include electron--phonon renormalization corrections.\cite{Miglio:2020p1,Antonius:2015p085137,Monserrat:2016p100301}
		\end{tablenotes}
	\end{threeparttable}
\end{table*}

\begin{figure}[t!]
	\includegraphics[width=3in]{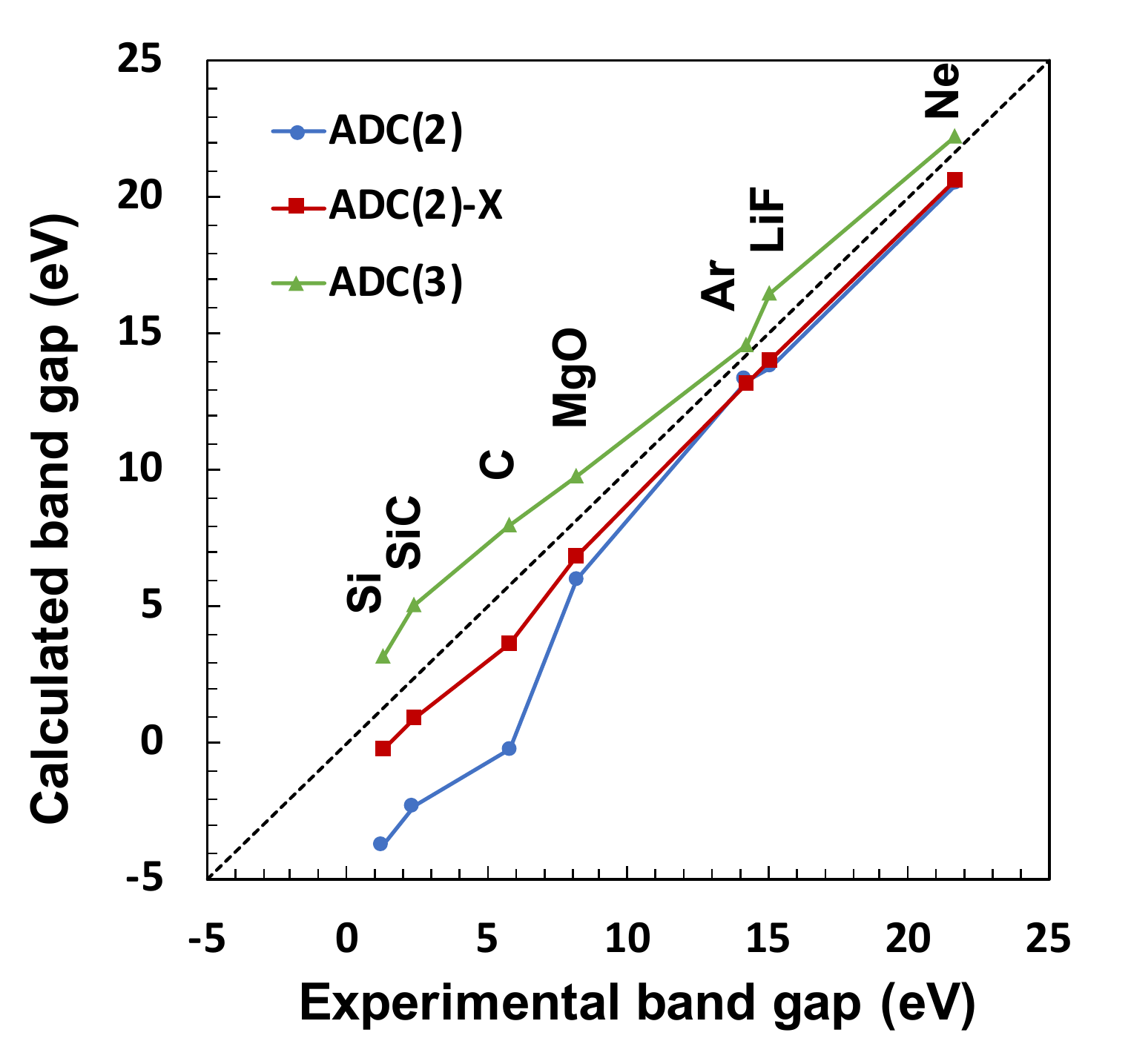}
	\captionsetup{justification=raggedright,singlelinecheck=false,font=footnotesize}
	\caption{Comparison of experimental fundamental band gaps with the band gaps computed using EA/IP-ADC(2), EA/IP-ADC(2)-X, and EA/IP-ADC(3) for seven semiconducting and insulating materials. 
		The experimental band gaps\cite{Madelung:2004semiconductors,Chiang:1989electronic,Schwentner:1975p528} include electron--phonon renormalization corrections.\cite{Miglio:2020p1,Antonius:2015p085137,Monserrat:2016p100301}
		Calculations used the gth-cc-pVTZ basis set and gth-hf-rev pseudopotential.
		The ADC(3) band gaps were estimated using \cref{eq:adc3_tz_gap}.
		See \cref{tab:band_gaps} for details. }
	\label{fig:band_gaps}
\end{figure} 

\subsection{Fundamental band gaps}
\label{sec:results:band_gaps}

Finally, we investigate the accuracy of EA/IP-ADC in predicting the fundamental band gaps of materials at the thermodynamic limit.
\cref{tab:band_gaps} reports the band gaps of Si, SiC, C, MgO, Ar, LiF, and Ne calculated using the EA/IP-ADC methods with the gth-cc-pV$X$Z ($X$ = D, T) basis sets and gth-hf-rev pseudopotential.\cite{Hong-Zhou:2022p1595}
For each material and basis set, the EA/IP-ADC band gaps were extrapolated to the thermodynamic limit as described in \cref{sec:computational_details}.
The results of EA/IP-ADC calculations with finite $k$-meshes are presented in the Supplementary Information.
\cref{tab:band_gaps} also shows the band gaps calculated using the periodic MP2 method with the all-electron cc-pVTZ basis set\cite{Lange:2021p081101} and the experimental band gaps corrected for the electron--phonon renormalization effects\cite{Madelung:2004semiconductors,Chiang:1989electronic,Schwentner:1975p528,Miglio:2020p1,Antonius:2015p085137,Monserrat:2016p100301,Lange:2021p081101} that range from 1.3 to 21.7 eV.

We first discuss how the computed fundamental band gaps depend on the size of the basis set.
For a fixed $k$-point mesh, the EA/IP-ADC band gaps show a weak basis set dependence with differences of less than 0.2 eV between the gth-cc-pVDZ and gth-cc-pVTZ results for all materials studied in this work.
The extrapolated band gaps reported in \cref{tab:band_gaps} show similar basis set dependence for EA/IP-ADC(2)-X and EA/IP-ADC(3). 
For EA/IP-ADC(2), larger (up to 0.4 eV) differences between the gth-cc-pVDZ and gth-cc-pVTZ results are observed for SiC, MgO, Ar, and LiF, which can be attributed to using different extrapolation schemes for these two basis sets (three-point vs two-point extrapolations, see \cref{sec:computational_details} for details).
Increasing the basis set lowers the band gap for all materials with the exception of Si and SiC, which show a small increase with the basis set size for some of the EA/IP-ADC methods.

We now turn our attention to \cref{fig:band_gaps}, which compares the experimental fundamental band gaps ($E_g^{exp}$) with the extrapolated band gaps computed using the EA/IP-ADC methods and the gth-cc-pVTZ basis set  ($E_g$).
Here, we use $E_g^{exp}$ as a reference and analyze how the errors in $E_g$ for each method change as the $E_g^{exp}$ becomes smaller from Ne to Si. 
For Ne, LiF, and Ar with $E_g^{exp}$ ranging from 21.7 to 14.2 eV, EA/IP-ADC(2) and EA/IP-ADC(2)-X show similar results underestimating the experimental band gaps by $0.9$ to $1.3$ eV.
In contrast, the EA/IP-ADC(3) methods overestimate $E_g^{exp}$ of these materials with errors ranging from 0.4 to 1.4 eV. 

As $E_g^{exp}$ becomes smaller, the results of EA/IP-ADC methods start to deviate from each other. 
In particular, EA/IP-ADC(2) severely underestimate $E_g$ of C, SiC, and Si (by 4.7 to 6 eV), predicting their values to be negative at the thermodynamic limit. 
These large errors of strict second-order ADC approximations are accompanied by the poor performance of the MP2 method, which underestimates $E_g$ of C, SiC, and Si by 3.4 to 4.2 eV.
Partially incorporating the third-order correlation effects in EA/IP-ADC(2)-X reduces the errors in $E_g$ for C, SiC, and Si by almost three-fold, from 4.7 -- 6 eV to 1.5 -- 2.1 eV.
As a result, the EA/IP-ADC(2)-X band gap errors show much weaker dependence on $E_g^{exp}$ than EA/IP-ADC(2), leading to a more parallel curve in \cref{fig:band_gaps}. 
Despite significant improvements, the EA/IP-ADC(2)-X methods still produce an unphysical negative $E_g$ for Si at the thermodynamic limit.
The EA/IP-ADC(3) methods overestimate the $E_g$ of C, SiC, and Si by 1.9 to 2.7 eV, predicting positive $E_g$ for all three materials. 

Overall, out of the three ADC approximations studied in this work, the $E_g$ computed using EA/IP-ADC(2)-X and EA/IP-ADC(3) show the best agreement with the available experimental data.
The EA/IP-ADC(2)-X results show the smallest mean absolute error (\mae = 1.38 eV) and standard deviation of errors (\std = 0.40 eV) relative to $E_g^{exp}$ and the most parallel curve in \cref{fig:band_gaps}.
The EA/IP-ADC(3) methods exhibit larger \mae (1.58 eV) and \std (0.82 eV), but correctly predict the sign of band gap for all materials.
Interestingly, we find that for each material the average of EA/IP-ADC(2)-X and EA/IP-ADC(3) band gaps computed using the gth-cc-pVTZ basis set is in a good agreement with $E_g^{exp}$ (\mae = 0.48 eV).
However, using a larger one-electron basis set is expected to lower the \mae of EA/IP-ADC(3) and increase the \mae of EA/IP-ADC(2)-X, shifting their average away from the experimental results. 

\section{Conclusions}
\label{sec:conclusions}

We presented the first implementation and benchmark of periodic non-Dyson algebraic diagrammatic construction theory for simulating charged excitations in three-dimensional solids (EA/IP-ADC). 
EA/IP-ADC allows to efficiently compute crystalline band structures, band gaps, and densities of states from the one-particle Green's functions approximated to low order in M\o ller--Plesset perturbation theory.
In contrast to the periodic Dyson algebraic diagrammatic construction theory reported earlier,\cite{Buth:2005p1,Buth:2006,Bezugly:2008p012006} the non-Dyson formalism\cite{Schirmer:1998p4734,Trofimov:2005p144115,Schirmer:2001p10621,Thiel:2003p2088,Dempwolff:2019p064108,Banerjee:2019p224112,Dempwolff:2020p024113,Banerjee:2021p074105,Dempwolff:2021p104117}  employed in our implementation allows for independent calculations of valence and conduction bands, which significantly reduces the computational cost.

Our work also features an implementation and benchmark of the second- and third-order periodic M\o ller--Plesset perturbation theory (MP2 and MP3), including the first-ever application of MP3 to three-dimensional crystalline systems.
The periodic MP2 and MP3 methods were used to calculate the equation-of-states and lattice parameters of diamond (C) and silicon (Si) crystals.
For both materials, adding the third-order electron correlation effects in MP3 significantly improves the accuracy of computed ground-state properties relative to periodic coupled cluster theory with single and double excitations (CCSD).

To assess the accuracy of periodic EA/IP-ADC for charged excited states of solids, we computed the band structures of C and Si using three EA/IP-ADC levels of theory (EA/IP-ADC(2), EA/IP-ADC(2)-X, and EA/IP-ADC(3)), a double-zeta Gaussian basis set, and a finite sampling of the Brillouin zone.
For C, the band structures calculated using EA/IP-ADC(2)-X and EA/IP-ADC(3) are in a very good agreement with a reference band structure from equation-of-motion CCSD (EOM-CCSD), while the EA/IP-ADC(2) band energies are uniformly underestimated by $\sim$ 2.3 eV.
For Si, all EA/IP-ADC methods show non-uniform errors in band energies in different points of the Brillouin zone, providing the evidence that small-gap semiconductors are challenging systems for finite-order perturbation theories such as EA/IP-ADC.

We benchmarked the accuracy of EA/IP-ADC further by calculating the thermodynamic-limit band gaps ($E_g$) of seven crystalline systems and compared them to the electron--phonon-corrected experimental band gaps ranging from 1.3 to 21.7 eV. 
For large-gap materials ($E_g$ $>$ 6 eV), the EA/IP-ADC methods combined with a triple-zeta basis set predict the band gaps with errors between 0.5 and 1.5 eV.
For materials with smaller experimental band gap, EA/IP-ADC(2) severely underestimate the band gap by $\sim$ 5-6 eV.
EA/IP-ADC(2)-X and EA/IP-ADC(3) significantly improve upon EA/IP-ADC(2) with errors ranging from 1.5 to 2.7 eV. 
EA/IP-ADC(2)-X shows the best performance overall, systematically underestimating the experimental band gap by $\sim$ 1.4 eV on average.
EA/IP-ADC(3) overestimates the experimental band gaps with a mean error of 1.6 eV, which is expected to become smaller when using a larger one-electron basis set.

The work presented herein represents an important first step towards making the non-Dyson algebraic diagrammatic construction theory routinely applicable to periodic systems.
Future work in our group will focus on a more efficient computer implementation of periodic EA/IP-ADC, developing new ADC approximations with improved accuracy, and extending periodic ADC to simulations of other excited-state properties of crystalline systems.

\suppinfo
See the supplementary material for the working equations of periodic EA/IP-ADC and the data used to calculate the EA/IP-ADC band gaps with finite $k$-meshes and at the thermodynamic limit.

\acknowledgement
	This work was supported by the start-up funds from the Ohio State University. 
	Additionally, S.B. was supported by a fellowship from Molecular Sciences Software Institute under NSF Grant No.\@ ACI-1547580.
	Computations were performed at the Ohio Supercomputer Center under projects PAS1583 and PAS1963.\cite{OhioSupercomputerCenter1987} 
	The authors would like to thank Garnet Chan, Timothy Berkelbach, Hong-Zhou Ye, and Xiao Wang for the discussions about linear dependencies in periodic Gaussian basis sets.
	The authors also thank Donna Odhiambo for assistance with some of the periodic ADC calculations.


\providecommand{\latin}[1]{#1}
\makeatletter
\providecommand{\doi}
  {\begingroup\let\do\@makeother\dospecials
  \catcode`\{=1 \catcode`\}=2 \doi@aux}
\providecommand{\doi@aux}[1]{\endgroup\texttt{#1}}
\makeatother
\providecommand*\mcitethebibliography{\thebibliography}
\csname @ifundefined\endcsname{endmcitethebibliography}
  {\let\endmcitethebibliography\endthebibliography}{}

\end{document}